\documentclass[12pt]{article}

\author{Yu.~M.~Zinoviev
       \thanks{E-mail address: Yurii.Zinoviev@ihep.ru} \\
        {\it Institute for High Energy Physics} \\
        {\it Protvino, Moscow Region, 142280, Russia}}
\title{All spin-2 cubic vertices \\
       with two derivatives}

\date{}

\begin{document}

\maketitle

\begin{abstract}
In this paper we provide a complete list of spin-2 cubic interaction
vertices with two derivatives. We work in (anti) de Sitter space with
dimension $d \ge 4$ and arbitrary value of cosmological constantе and
use simple metric formalism without any auxiliary or Stueckelberg
fields. We separately consider cases with one, two and three different
spin-2 fields entering the vertex where each field may be massive,
massless or partially massless one. The connection of our results with
massive (bi)gravity theories is also briefly discussed.
\end{abstract}

\thispagestyle{empty}
\newpage
\setcounter{page}{1}

\section*{Introduction}

One of the effective ways for investigation of possible higher spin
interactions is the so called constructive approach. Here one assumes
that the whole Lagrangian of the theory can be considered as a row
$$
{\cal L} = {\cal L}_0 + {\cal L}_1 + {\cal L}_2 + \dots
$$
where ${\cal L}_0$ --- free Lagrangian, ${\cal L}_1$ contains all
cubic vertices and so on. Similarly, for gauge transformations one
also assumes
$$
\delta = \delta_0 + \delta_1 + \delta_2 + \dots
$$
where $\delta_0$ --- in-homogeneous terms, corresponding to gauge
invariance of free Lagrangian, $\delta_1$ is linear in fields and so
on. The first and very important step in all such investigations is
the so called linear approximation, i.e. cubic vertices in the
Lagrangian and linear in fields (hence the name) terms in gauge
transformations. In this approximation gauge invariance requires that
$$
\delta_0 {\cal L}_1 + \delta_1 {\cal L}_0 = 0
$$
and it means that variation of the cubic vertex under the free gauge
transformations must be proportional to free equations and thus can be
compensated by appropriate corrections to gauge transformations. The
importance of linear approximation comes from the fact that due to
linearity of gauge transformations the properties of cubic vertex for
any three given fields do not depend on the presence or absence of any
other fields in the system. Thus to a large extent the results of
linear approximation are model independent and this open the
possibility for classification of all consistent cubic vertices.

For the last 10-15 years a lot of interesting and important results
for the cubic vertices where obtained using very different approaches
(light cone or Lorentz covariant, metric-like or frame-like)
\cite{Vas01}-\cite{ST10}.\footnote{Certainly, this list of References
is not in any way complete, it serves just as an illustration of very
different approaches used.} In spite of very simple formulation any
real investigations of cubic vertices require very complicated and
cumbersome calculations. To simplify this task the so called 
TT-approach was introduced \cite{JT11,JLT12,JLT12a,JLT12b} where one
try to construct the core part of cubic vertex that survives when all
the fields entering the vertex are subject to transversality and
tracelessness constraints. Such approach nicely work for the massless
fields where gauge invariance allows one to reconstruct full vertex
with the TT-constraints relaxed. But when massive or partially
massless fields are present it is not at all evident if such TT-vertex
(or some particular combination of them) can be uplifted to the full
unconstrained one. 

The aim of this paper --- carefully consider may be the most simple
case, namely all cubic vertices for spin-2 fields with two
derivatives.\footnote{Even this simple case requires rather long
calculations so that one must use some computer algebra system. In
this work we used Reduce \cite{Red}.} Besides being an illustrative
example for the construction of full cubic vertices, our results also
have some relations with recent investigations of massive gravity
\cite{RG10,RGT10,HRS11}, bigravity \cite{HR11,HR11b,HSS12b} and
multigravity \cite{HR12,HSS12a} theories as well as with attempts to
construct partially massless gravity or bigravity
\cite{RR12,HSS12,HSS12d,DJW12,DSW13,RHRT13}. We work in a simple
metric formalism where spin-2 field is described by symmetric second
rank tensor $h_{\mu\nu}$ without any auxiliary or Stueckelberg fields.
To be sure that we do not miss some particular cases we separately
consider vertices with one, two and three different spin-2 fields, in
this, any field can be massive, massless or partially massless one. We
use "brute force" method, i.e. we construct the most general
expression for cubic vertex and require that both transversality and
tracelessness constraints follow from Lagrangean equations. Let us
present here the main results of our work leaving technical details
for the main text.

\noindent
{\bf One field}

\begin{itemize}
\item {\bf Massless case}. There is a unique (up to possible field
redefinitions) solution. Our choice corresponds to gauge invariance
under
$$
\delta h_{\mu\nu} = D_{(\mu} \xi_{\nu)} + a_0 ( \xi^\alpha D_\alpha
h_{\mu\nu} + D_{(\mu} \xi^\alpha h_{\nu)\alpha} )
$$
\item {\bf Massive case}. General solution has the cubic potential of
the form:
$$
{\cal L}_{10} = \frac{a_0 m^2}{4} [ (1+2b_3) h_{\mu\nu} h_{\nu\alpha}
h_{\alpha\mu} - (1+3b_3) h h_{\mu\nu}{}^2 + b_3 h^3 ]
$$
In this scalar constraint has the structure:
$$
{\cal C} \sim h \oplus Dh Dh \oplus h^2
$$
Terms with derivatives are absent for $b_3 = \frac{1}{2}$, in this
$$
{\cal L}_{10} = \frac{a_0m^2}{2} [ h_{\mu\nu} h_{\nu\alpha}
h_{\alpha\mu} - \frac{5}{4} h h_{\mu\nu}{}^2 + \frac{1}{4} h^3 ]
$$
\item {\bf Partially massless case $m^2 = (d-2) \kappa$}. Only
solution with $b_3 = \frac{1}{2}$ admit partially massless limit in de
Sitter space but for $d=4$ only!
\end{itemize}

\noindent
{\bf Two fields}. Here we call "first field" the one that enters the
vertex linearly while "second field" is the one entering it
quadratically.
\begin{itemize}
\item {\bf Both fields are massless}. There is a unique (up to
possible field redefinitions) solution. Our choice corresponds to
gauge invariance under
\begin{eqnarray*}
\delta h_{1\mu\nu} &=& D_{(\mu} \xi_{1\nu)} + a_0 ( \xi_2{}^\alpha
D_\alpha h_{2\mu\nu} + D_{(\mu} \xi_2{}^\alpha h_{2\nu)\alpha} ) \\
\delta h_{2\mu\nu} &=& D_{(\mu} \xi_{2\nu)} + a_0 ( \xi_1{}^\alpha
h_{2\mu\nu} + D_{(\mu} \xi_1{}^\alpha h_{2\nu)\alpha} + \xi_2{}^\alpha
D_\alpha h_{1\mu\nu} + D_{(\mu} \xi_2{}^\alpha h_{1\nu)\alpha} )
\end{eqnarray*}
\item {\bf Both fields are massive}. General solution has cubic
potential of the form:
\begin{eqnarray*}
{\cal L}_{01} &=& \frac{a_0}{4} [ (m_1{}^2 + 2m_2{}^2 + 2b_4)
h_{1\mu\nu} h_{2\nu\alpha} h_{2\alpha\mu} - (m_1{}^2 + b_4) h_1
h_{2\mu\nu}{}^2 \\
 && \quad - 2(m_2{}^2 + b_4) h_{1\mu\nu} h_2 h_{2\mu\nu} + b_4 h_1
h_2{}^2 ]
\end{eqnarray*}
 Scalar constraints for both fields are algebraic only for
$$
b_4 = \frac{m_1{}^2 + 2m_2{}^2}{2}
$$
In this:
\begin{eqnarray*}
{\cal L}_{01} &=& \frac{a_0}{2} [ (m_1{}^2 + 2m_2{}^2) h_{1\mu\nu}
h_{2\nu\alpha} h_{2\alpha\mu} - \frac{3m_1{}^2 + 2m_2{}^2}{4} h_1
h_{2\mu\nu}{}^2 \\
 && \quad - \frac{m_1{}^2 + 4m_2{}^2}{2} h_{1\mu\nu} h_2 h_{2\mu\nu} +
\frac{m_1{}^2 + 2m_2{}^2}{4} h_1 h_2{}^2 ]
\end{eqnarray*}
\item {\bf First field is massless $m_1 = 0$}. There is a unique
solution:
$$
{\cal L}_{01} = a_0 m_2{}^2 [ h_{1\mu\nu} h_{2\nu\alpha}
h_{2\alpha\mu} - \frac{1}{4} h_1 h_{2\mu\nu}{}^2 - h_{1\mu\nu} h_2
h_{2\mu\nu} + \frac{1}{4} h_1 h_2{}^2 ]
$$
Scalar constraint for second field is algebraic. In $dS$ this solution
admits partially massless limit $m_2{}^2 \to (d-2)\kappa$ without
restrictions on $d$.
\item {\bf Second field is massless $m_2 = 0$}. There is no solution
except $m_1 = 0$.
\item {\bf First field is partially massless $m_1{}^2 = (d-2)\kappa$}.
Solution exists only for
$$
m_2{}^2 = \frac{d(d-2)\kappa}{4}
$$
Scalar constraint for the second field is algebraic. Note that for
$d=4$ this corresponds to partially massless case.
\item {\bf Second field is partially massless $m_2{}^2 =
(d-2)\kappa$}. Solution exists only for
$$
m_1{}^2 = 2(d-3) \kappa
$$
Scalar constraint for the first field is algebraic. Note that for
$d=4$ this again corresponds to partially massless case.
\end{itemize}

\noindent
{\bf Three fields}.

\begin{itemize}
\item {\bf All three fields are massless}. There is a unique (up to
possible fields redefinitions) solution. Our choice corresponds to
gauge invariance under
\begin{eqnarray*}
\delta h_{1\mu\nu} &=& D_{(\mu} \xi_{1\nu)} + a_0 ( \xi_2{}^\alpha
D_\alpha h_{3\mu\nu} + D_{(\mu} \xi_2{}^\alpha h_{3\nu)\alpha} +
\xi_3{}^\alpha h_{2\mu\nu} + D_{(\mu} \xi_3{}^\alpha h_{2\nu)\alpha} )
\\
\delta h_{2\mu\nu} &=& D_{(\mu} \xi_{2\nu)} + a_0 ( \xi_1{}^\alpha
D_\alpha h_{3\mu\nu} + D_{(\mu} \xi_1{}^\alpha h_{3\nu)\alpha} +
\xi_3{}^\alpha h_{1\mu\nu} + D_{(\mu} \xi_3{}^\alpha h_{1\nu)\alpha} )
\\
\delta h_{3\mu\nu} &=& D_{(\mu} \xi_{3\nu)} + a_0 ( \xi_1{}^\alpha
D_\alpha h_{2\mu\nu} + D_{(\mu} \xi_1{}^\alpha h_{2\nu)\alpha} +
\xi_2{}^\alpha h_{1\mu\nu} + D_{(\mu} \xi_2{}^\alpha h_{1\nu)\alpha} )
\end{eqnarray*}
\item {\bf Two fields are massless $m_1 = m_2 = 0$}. No solution
except $m_3 = 0$.
\item {\bf One field is massless $m_1 = 0$}. Solution exists for $m_2
= m_3$ only. Scalar constraints are algebraic. In $dS$ this solution
admits partially massless limit.
\item {\bf All three fields are massive}. General solution has
potential
\begin{eqnarray*}
{\cal L}_{10} &=& \frac{a_0}{2} [ (m_1{}^2 + m_2{}^2 + m_3{}^2 + 2b_5)
h_{1\mu\nu} h_{2\nu\alpha} h_{3\alpha\mu} - (m_3{}^2 + b_5)
h_{1\mu\nu} h_{2\mu\nu} h_3 \\
 && \quad - (m_2{}^2 + b_5) h_{1\mu\nu} h_2 h_{3\mu\nu} - (m_3{}^2 +
b_5) h_1 h_{2\mu\nu} h_{3\mu\nu} + b_5 h_1 h_2 h_3 ]
\end{eqnarray*}
Terms with derivatives in scalar constraints are absent for
$$
b_5 = \frac{m_1{}^2 + m_2{}^2 + m_3{}^2}{2}
$$
In this
\begin{eqnarray*}
{\cal L}_{10} &=& a_0 [ (m_1{}^2 + m_2{}^2 + m_3{}^2)
h_{1\mu\nu} h_{2\nu\alpha} h_{3\alpha\mu} - \frac{m_1{}^2 + m_2{}^2 +
3m_3{}^2}{4} h_{1\mu\nu} h_{2\mu\nu} h_3 \\
 && \quad - \frac{m_1{}^2 + 3m_2{}^2 + m_3{}^2}{4} h_{1\mu\nu} h_2
h_{3\mu\nu} - \frac{3m_1{}^2 + m_2{}^2 + m_3{}^2}{4} h_1 h_{2\mu\nu}
h_{3\mu\nu}  \\
 && \quad + \frac{m_1{}^2 + m_2{}^2 + m_3{}^2}{4} h_1 h_2 h_3 ]
\end{eqnarray*}
\item {\bf One field is partially massless $m_1{}^2 = (d-2)\kappa$}.
There is a solution provided
$$
(m_2{}^2 - m_3{}^2)^2 + 2(m_2{}^2 + m_3{}^2) \kappa = d(d-2) \kappa^2
$$
In this scalar constraints for both massive fields are algebraic.
There are two particular solutions of the last relation that
correspond to some two field cases given above.
\begin{itemize}
\item If masses of the second and third fields are equal $m_2 = m_3$
then this relation gives
$$
m_2{}^2 = m_3{}^2 = \frac{d(d-2)\kappa}{4}
$$
and this corresponds to the two field case where first field is
partially massless.
\item If the second field is also partially massless $m_2{}^2 =
(d-2)\kappa$ then we obtain
$$
m_3{}^2 = 2(d-3)\kappa
$$
exactly as in the two field case where the second field is partially
massless.
\end{itemize}
\end{itemize}

The layout of the paper is simple and straightforward. In Section 1 we
provide all necessary kinematic formulas as well as our conventions.
Section 2 devoted to the case of single spin-2 field. Surely, this
case is rather well understood by now but it is instructive to
reproduce these results by the same method that subsequently will be
used for the case with two and three spin-2 fields. In Section 3 we
consider two field case, while in Section 4 we consider relation of
our results with formulation of bigravity \cite{HR11,HR11b,HSS12b}. At
last, in Section 5 we consider cubic vertices for three fields with
different masses.

\section{Kinematics}

We will work in (anti) de Sitter space-time with dimension $d \ge 4$
and arbitrary value of cosmological constant $\Lambda$. Indices are
raised and lowered with non-dynamical background metric
$\eta_{\mu\nu}$, while $(A)dS$ covariant derivative $D_\mu$ is
normalized so that
\begin{equation}
[ D_\mu, D_\nu ] \xi_\alpha = - \kappa (\eta_{\mu\alpha} \xi_\nu -
\eta_{\nu\alpha} \xi_\mu), \qquad \kappa = \frac{2\Lambda}{(d-1)(d-2)}
\end{equation}
We use metric like formalism where spin-2 field is described by
symmetric second rank tensor $h_{\mu\nu}$ and choose free Lagrangian
for massive field in the form:\footnote{Note that due to
non-commutativity of $(A)dS$ covariant derivatives there is an
ambiguity in the choice of combination of the second and third terms,
in this the structure of the terms without derivatives depends on the
choice made.}
\begin{eqnarray}
{\cal L}_0 &=& \frac{1}{2} [ D_\alpha h_{\mu\nu} D_\alpha h_{\mu\nu} -
D_\mu h_{\nu\alpha} D_\nu h_{\mu\alpha} - (Dh)_\mu (Dh)_\mu + 2
(Dh)_\mu D_\mu h - D_\mu h D_\mu h ] \nonumber \\
 && - \frac{m^2 - \kappa(d-2)}{2} [ h_{\mu\nu}{}^2 - h^2 ]
\end{eqnarray}
where $(Dh)_\mu = D^\nu h_{\mu\nu}$ and $h = h_\mu{}^\mu$.

As is well known the correct number of physical degrees of freedom
requires that two constraints\footnote{Here and in what follows
dynamical equations will be of second order in derivatives, thus any
relation containing field and its first derivative only will be
considered as constraint.} --- vector and scalar ones --- follow from
the Lagrangean equations. It is easy to check that for the free
Lagrangian given above we indeed have them:
\begin{eqnarray}
{\cal C}_\nu &=& D^\mu \frac{\delta {\cal L}_0}{\delta h_{\mu\nu}} = -
m^2 ((Dh)_\nu - D_\nu h) = 0 \\
{\cal C} &=& (D^\mu D^\nu - \frac{m^2}{(d-2)} \eta^{\mu\nu}) 
\frac{\delta {\cal L}_0}{\delta h_{\mu\nu}} = - \frac{(d-1)}{(d-2)}
m^2 [ m^2 - \kappa(d-2)] h = 0
\end{eqnarray}
From these relations one can immediately note that there are two
special cases. The first case is the massless one $m = 0$ where we
loose both constraints but the Lagrangian instead becomes invariant
under the local gauge transformations with vector parameter:
\begin{equation}
\delta_0 h_{\mu\nu} = D_\mu \xi_\nu + D_\nu \xi_\mu
\end{equation}
The second case is the so called partially massless one 
\cite{DW01,DW01a,DW01c,Zin01} $m^2 = \kappa(d-2)$, in this we loose
the second constraint but the Lagrangian becomes invariant under the
local gauge transformations with scalar parameter:
\begin{equation}
\delta_0 h_{\mu\nu} = (D_\mu D_\nu - \frac{m^2}{(d-2)} \eta_{\mu\nu})
\xi
\end{equation}
Note that to have correct number of physical degrees of freedom it is
important that we still have vector constraint.

\section{One field}

In this Section we consider cubic vertex with one spin-2 field, i.e.
its self-interaction. By now this case is rather well understood but
it is instructive to reproduce known results by the same method that
will subsequently be used for vertices with two and three different
spin-2 fields.

We will look for the cubic vertex in the following form:
$$
{\cal L}_1 = {\cal L}_{12} + {\cal L}_{10}
$$
where ${\cal L}_{12}$ contains terms with two derivatives which without
loss of generality can be chosen in the form:\footnote{Similarly to
the free case due to non-commutativity of $(A)dS$ covariant
derivatives there are some ambiguities in the choice of expression for
${\cal L}_{12}$ and this choice determines the explicit dependence for
the coefficients in ${\cal L}_{10}$ on the cosmological constant. In
this, all physical results do not depend on the choice made.}
$$
{\cal L}_{12} \sim h D h D h
$$
while ${\cal L}_{10}$ contains terms without derivatives and looks
like:
\begin{equation}
{\cal L}_{10} = b_1 h_{\mu\nu} h_{\nu\alpha} h_{\alpha\mu} + b_2 h
h_{\mu\nu}{}^2 + b_3 h^3
\end{equation}

First of all we require that after switching on interaction we will
still have vector constraint but in general with some corrections
quadratic in field:
$$
\Delta {\cal C}_\nu \approx D^\mu \frac{\delta {\cal L}_1}{\delta
h_{\mu\nu}}
$$
where here and in what follows $\approx$ means "on the free mass
shell", i.e. up to the terms proportional to free equations.
Schematically this can be written as
$$
\Delta {\cal C}_V = D \frac{\delta {\cal L}_1}{\delta h} + (Dh + h D)
\frac{\delta {\cal L}_0}{\delta h}
$$
where $(Dh + h D)$ denotes the most general operator linear in field
$h$ and of first order in derivatives. It turns out that the general
solution for ${\cal L}_{12}$ has five free parameters. Recall that in
any case where interacting vertex has the same (or higher) number of
derivatives as the free Lagrangian one always faces the ambiguities
related with the possibility to make field redefinitions and hence
obtain the families of physically equivalent theories. In the case at
hands possible field redefinitions have the form:
\begin{equation}
h_{\mu\nu} \Rightarrow h_{\mu\nu} + s_1 h_{\mu\alpha} h_{\alpha\nu} +
s_2 h h_{\mu\nu} + s_3 \eta_{\mu\nu} h_{\alpha\beta}{}^2 + s_4
\eta_{\mu\nu} h^2
\end{equation}
thus leaving us with only one non-trivial coupling constant. Using
this freedom it is always possible to choose the form of 
${\cal L}_{12}$ such that in the massless case $m=0$ the Lagrangian
will be invariant under the following gauge transformations:
\begin{equation}
\delta h_{\mu\nu} = D_\mu \xi_\nu + D_\nu \xi_\mu + a_0 [ \xi^\alpha
D_\alpha h_{\mu\nu} + D_\mu \xi^\alpha h_{\alpha\nu} + D_\nu
\xi^\alpha h_{\mu\alpha} ]
\end{equation}

Now let us turn to the scalar constraint. Here we also require that we
will still have this constraint with possible corrections quadratic in
field $h$:
$$
\Delta {\cal C} \approx (D^\mu D^\nu - \frac{m^2}{(d-2)}
\eta^{\mu\nu}) \frac{\delta {\cal L}_1}{\delta h_{\mu\nu}}
$$
or schematically
$$
\Delta {\cal C} = (D D - \eta) \frac{{\cal L}_1}{\delta h} + 
(D^2h + Dh D + h D^2 + h) \frac{{\cal L}_0}{\delta h}
$$
where $(D^2h + Dh D + h D^2 + h)$ denotes the most general operator
linear in field $h$ and of second order in derivatives. General
solution has the form:
\begin{equation}
b_1 = \frac{m^2 a_0 + 8b_3}{4}, \qquad
b_2 = - \frac{m^2 a_0 + 12b_3}{4}
\end{equation}
in agreement with the results of \cite{RG10}. However, for general
values of parameter $b_3$ the scalar constraint has the following
structure:
$$
{\cal C} \sim h + Dh Dh + h^2
$$
so it contains terms with first derivatives of $h$. But such terms
lead to the problems with causality \cite{DW12,DSW13}. Happily, there
is a special value of this parameter:
$$
b_3 = \frac{m^2a_0}{8}
$$
when all these terms are absent so that scalar constraint remains to
be purely algebraic, in this
\begin{equation}
{\cal L}_{10} = \frac{m^2a_0}{2} [ h_{\mu\nu} h_{\nu\alpha}
h_{\alpha\mu} - \frac{5}{4} h h_{\mu\nu}{}^2 +  \frac{1}{4} h^3] 
\end{equation}
Note that this is exactly the same structure that was obtained in
\cite{Zin06} using gauge invariant description of massive spin-2
field. Moreover it is this solution that in de Sitter space admits
partially massless limit in $d=4$ \cite{Zin06,DJW13,RHRT13}. In this
the Lagrangian is invariant under the following gauge transformations
(compare with Eq. (2.44) in \cite{RHRT13}\footnote{In our conventions
round brackets denote symmetrization without normalization factor and
this may explain slightly different coefficients.}):
\begin{eqnarray}
\delta h_{\mu\nu} &=& (D_\mu D_\nu - \frac{m^2}{2} \eta_{\mu\nu} ) \xi
+ \tilde{a}_0 (D_\mu D_\nu - \frac{m^2}{2} \eta_{\mu\nu} ) (h\xi)
\nonumber \\
 && + \frac{a_0}{4} [ h_{(\mu}{}^\alpha D_{\nu)} D_\alpha \xi -
D_{(\mu} h_{\nu)}{}^\alpha D_\alpha \xi + 2 D^\alpha h_{\mu\nu}
D_\alpha \xi - m^2 h_{\mu\nu} \xi ]
\end{eqnarray}
where the terms with the factor $\tilde{a}_0$ correspond to
redefinition of gauge parameter $\xi$ and could be discarded.

\section{Two fields}

In this Section we consider cubic vertex for two spin-2 fields with
different masses $m_1$ and $m_2$, in this we will assume that first
field $h_{1\mu\nu}$ enters vertex linearly and the second one
$h_{2\mu\nu}$ ---- quadratically. Without loss of generality the part
of the vertex with two derivatives can be chosen in the form (again
with a lot of ambiguities due to non-commutativity of covariant
derivatives):
$$
{\cal L}_{12} \sim h_1 h_2 D^2 h_2 \oplus h_1 D h_2 D h_2 
$$
while terms without derivatives will be written as follows:
\begin{equation}
{\cal L}_{10} = b_1 h_{1\mu\nu} h_{2\nu\alpha} h_{2\alpha\mu} + b_2
h_1 h_{2\mu\nu}{}^2 + b_3 h_2 h_{1\mu\nu} h_{2\mu\nu} + b_4 h_1
h_2{}^2
\end{equation}

First of all we require that we still have both vector constraints
with optional quadratic corrections:
$$
\Delta {\cal C}_{1\nu} \approx D^\mu \frac{\delta {\cal L}_1}{\delta
h_{1\mu\nu}}, \qquad \Delta {\cal C}_{2\nu} \approx D^\mu 
\frac{\delta {\cal L}_1}{\delta h_{2\mu\nu}}
$$
or schematically:
\begin{eqnarray*}
\Delta {\cal C}_{1V} = D \frac{\delta {\cal L}_1}{\delta h_1} + (D h_2
+ h_2 D) \frac{\delta {\cal L}_0}{\delta h_2} \\
\Delta {\cal C}_{2V} = D \frac{\delta {\cal L}_1}{\delta h_2} + (D h_1
+ h_1 D) \frac{\delta {\cal L}_0}{\delta h_2}  + (D h_2 + h_2 D)
\frac{\delta {\cal L}_0}{\delta h_1}
\end{eqnarray*}
It turns out that general solution for ${\cal L}_{12}$ has ten free
parameters, but taking into account possible field redefinitions that
in this case look like
\begin{eqnarray}
h_{1\mu\nu} &\Rightarrow& h_{1\mu\nu} + s_1 h_{2\mu\alpha}
h_{2\alpha\nu} + s_2 h_2 h_{2\mu\nu} + \eta_{\mu\nu} ( s_3
h_{2\alpha\beta}{}^2 + s_4 h_2{}^2)  \\
h_{2\mu\nu} &\Rightarrow& h_{2\mu\nu} + s_5 h_{1\alpha(\mu}
h_{2\nu)\alpha} + s_6 h_1 h_{2\mu\nu} + s_7 h_2 h_{1\mu\nu} + 
\eta_{\mu\nu} (s_8 h_{1\alpha\beta} h_{2\alpha\beta} + s_9 h_1 h_2)
\nonumber
\end{eqnarray}
we again see that there is only one non-trivial coupling constant
here. Using this freedom one can always bring ${\cal L}_{12}$ into the
form such that in the massless case $m_1 = m_2 = 0$ the Lagrangian
will be invariant under the following local gauge transformations:
\begin{eqnarray}
\delta h_{1\mu\nu} &=& D_\mu \xi_{1\nu} + D_\nu \xi_{1\mu} + a_0 [
\xi_2{}^\alpha D_\alpha h_{2\mu\nu} + D_{(\mu} \xi_2{}^\alpha
h_{2\nu)\alpha} ]  \\
\delta h_{2\mu\nu} &=& D_\mu \xi_{2\nu} + D_\nu \xi_{2\mu} + a_0 [
\xi_2{}^\alpha D_\alpha h_{1\mu\nu} + D_{(\mu} \xi_2{}^\alpha
h_{1\nu)\alpha} + \xi_1{}^\alpha D_\alpha h_{2\mu\nu} + D_{(\mu}
\xi_1{}^\alpha h_{2\nu)\alpha} ] \nonumber
\end{eqnarray}

Now let us turn to the scalar constraints with possible quadratic
corrections:
$$
\Delta {\cal C}_1 \approx (D^\mu D^\nu - \frac{m_1{}^2}{(d-2)}
\eta^{\mu\nu}) \frac{\delta {\cal L}_1}{\delta h_{1\mu\nu}}, \qquad
\Delta {\cal C}_2 \approx (D^\mu D^\nu - \frac{m_2{}^2}{(d-2)}
\eta^{\mu\nu}) \frac{\delta {\cal L}_1}{\delta h_{2\mu\nu}}
$$
or schematically:
\begin{eqnarray*}
\Delta {\cal C}_1 &=& (D D - \eta) \frac{{\cal L}_1}{\delta h_1} + 
(D^2h_2 + Dh_2 D + h_2 D^2 + h_2) \frac{{\cal L}_0}{\delta h_2} \\
\Delta {\cal C}_2 &=& (D D - \eta) \frac{{\cal L}_1}{\delta h_2} + 
(D^2h_1 + Dh_1 D + h_1 D^2 + h_1) \frac{{\cal L}_0}{\delta h_2}  \\
 && + (D^2h_2 + Dh_2 D + h_2 D^2 + h_2) \frac{{\cal L}_0}{\delta h_1}
\end{eqnarray*}

In this case we have a number of possibilities because each spin-2
field can be massive, massless or partially massless one. In what
follows we consider all of them one by one.

\noindent
{\bf Both fields are massive}. General solution has the form:
\begin{equation}
b_1 = \frac{a_0(m_1{}^2 + 2m_2{}^2) + 8b_4}{4}, \qquad
b_2 = - \frac{a_0m_1{}^2 + 4b_4}{4}, \qquad
b_3 = - \frac{a_0 m_2{}^2 + 4b_4}{2}
\end{equation}
so we have one free parameter $b_4$. But as in the previous case for
general values of this parameter both scalar constraints contain
dangerous terms with derivatives:
\begin{eqnarray*}
{\cal C}_1 &\sim& h_1 + Dh_2 Dh_2 + h_2{}^2 \\
{\cal C}_2 &\sim& h_2 + Dh_1 Dh_2 + h_1 h_2
\end{eqnarray*}
But there is a special value
\begin{equation}
b_4 = \frac{a_0(m_1{}^2 + 2m_2{}^2)}{8}
\end{equation}
when all these terms are absent, in this
\begin{equation}
b_1 = \frac{a_0(m_1{}^2+2m_2{}^2)}{2}, \qquad
b_2 = - \frac{a_0(3m_1{}^2+2m_2{}^2)}{8}, \qquad
b_3 = - \frac{a_0(m_1{}^2+4m_2{}^2)}{4}
\end{equation}

\noindent
{\bf First field is massless $m_1 = 0$}. Here there exists unique
solution:
\begin{equation}
b_1 = - b_3 = a_0 m_2{}^2, \qquad
b_2 = - b_4 = - \frac{a_0 m_2{}^2}{4}
\end{equation}
corresponding to usual gravitational interaction for massive spin-2
particle. Scalar constraint for massive field appears to be purely
algebraic without any derivative terms. Moreover in de Sitter space
this solution admits partially massless limit without any restrictions
on the dimension $d$.

\noindent
{\bf Second field is massless $m_2=0$}. There is no solution here
(except for the $m_1 = 0$) in agreement with the results of
\cite{Met12} that such vertex requires as many as six derivatives.

\noindent
{\bf First field is partially massless $m_1{}^2 = \kappa(d-2)$}. There
exists solution only for
\begin{equation}
m_2{}^2 = \frac{d(d-2)\kappa}{4}
\end{equation}
Second scalar constraint is algebraic and the Lagrangian is invariant
under the following gauge transformations:
\begin{eqnarray}
\delta h_{1\mu\nu} &=& (D_\mu D_\nu - \frac{m_1{}^2}{2} \eta_{\mu\nu}
) \xi \\
\delta h_{2\mu\nu} &=& \frac{a_0}{2} [ \frac{(d-2)}{d} 
h_{2(\mu}{}^\alpha D_{\nu)} D_\alpha \xi - \frac{2}{d} D_{(\mu}
h_{2\nu)}{}^\alpha D_\alpha \xi + D^\alpha h_{2\mu\nu} D_\alpha \xi +
\frac{(d-6)\kappa}{2} h_{2\mu\nu} \xi ] \nonumber
\end{eqnarray}
Note that for $d=4$ this corresponds to partially massless case
$m_2{}^2 = 2\kappa$.

\noindent
{\bf Second field is partially massless $m_2{}^2 = \kappa(d-2)$}.
There exists solution only for
\begin{equation}
m_1{}^2 = 2(d-3)\kappa
\end{equation}
First scalar constraint is algebraic and the Lagrangian is invariant
under the following gauge transformations:
\begin{eqnarray}
\delta h_{1\mu\nu} &=& \frac{a_0}{4} [ h_{2(\mu}{}^\alpha D_{\nu)}
D_\alpha \xi - D_{(\mu} h_{2\nu)}{}^\alpha D_\alpha \xi + 2 D^\alpha
h_{2\mu\nu} D_\alpha \xi -  2\kappa h_{2\mu\nu} \xi ] \nonumber \\
\delta h_{2\mu\nu} &=& (D_\mu D_\nu - \kappa \eta_{\mu\nu}) \xi +
\tilde{a}_0 (D_\mu D_\nu - \kappa \eta_{\mu\nu}) (h_1\xi) \\
 && + \frac{a_0}{2} [ \frac{(d-3)}{(d-2)} h_{1(\mu}{}^\alpha D_{\nu)}
D_\alpha \xi - \frac{1}{(d-2)} D_{(\mu} h_{1\nu)}{}^\alpha D_\alpha
\xi + D^\alpha h_{1\mu\nu} D_\alpha \xi + \kappa(d-5) h_{1\mu\nu} \xi
] \nonumber
\end{eqnarray} 
For $d=4$ this again corresponds to partially massless case $m_1{}^2 =
2\kappa$.

\section{Bigravity}

As is well known (see e.g. \cite{BDGH00}) any theory for two massless
spin-2 fields with no more than two derivatives (and when both fields
are physical one and not ghost) by field redefinitions can be brought
into the form with two independent parts, each one being just usual
gravity theory for some metric:
$$
{\cal L} = {\cal L}(g_{\mu\nu}) + {\cal L}(f_{\mu\nu})
$$
It is these separated metrics that where used in formulation of so
called bigravity theory \cite{HR11,HR11b,HSS12b} where mixing appears
in the potential terms only. The aim of this Section is to show how
these two metrics can be related with massless and massive spin-2
fields.

Let us denote $h_0$ the field that remains to be massless and $h_m$
--- the one that becomes massive. There are four possible cubic
vertices for two massless spin-2 fields:
$$
{\cal L}_1 \sim a_{01} h_0{}^3 \oplus a_{02} h_0{}^2 h_m \oplus a_{03}
h_0 h_m{}^2 \oplus a_{04} h_m{}^3
$$
which at this level are completely independent. But we already know
that there is no solution for cubic vertex where massive field enters
linearly, thus we put $a_{02} = 0$. Then collecting the results from
previous two sections for gauge transformations we will have:
\begin{eqnarray}
\delta h_{0\mu\nu} &=& D_\mu \xi_{0\nu} + D_\nu \xi_{0\mu} +
a_{01} [ \xi_0{}^\alpha D_\alpha h_{0\mu\nu} + D_{(\mu} \xi_0{}^\alpha
h_{0\nu)\alpha} ] \nonumber \\
 && + a_{03} [ \xi_m{}^\alpha D_\alpha h_{m\mu\nu} + D_{(\mu}
\xi_m{}^\alpha h_{m\nu)\alpha} ] \nonumber \\
\delta h_{m\mu\nu} &=& D_\mu \xi_{m\nu} + D_\nu \xi_{m\mu} 
 + a_{03} [ \xi_0{}^\alpha D_\alpha h_{m\mu\nu} + D_{(\mu}
\xi_0{}^\alpha h_{m\nu)\alpha} ] \\
 && + a_{03} [ \xi_m{}^\alpha D_\alpha h_{0\mu\nu} + D_{(\mu}
\xi_m{}^\alpha h_{0\nu)\alpha} ] \nonumber \\
 && + a_{04} [ \xi_m{}^\alpha D_\alpha h_{m\mu\nu} + D_{(\mu}
\xi_m{}^\alpha h_{m\nu)\alpha} ] \nonumber
\end{eqnarray}
A nice property of this form for gauge transformations is that their
algebra can be closed without any corrections beyond linear
approximation. For the case at hands this requires $a_{01} =
a_{03}$,\footnote{Note that this relation is nothing else but usual
manifestation of universality of gravitational interactions.} while
$a_{04}$ can be arbitrary. But in this case if we make a change of
variables:
\begin{equation}
h_1 = h_0 \cos(\theta) + h_m \sin(\theta), \qquad
h_2 = - h_0 \sin(\theta) + h_m \cos(\theta)
\end{equation}
and similarly
\begin{equation}
\xi_1 = \xi_0 \cos(\theta) + \xi_m \sin(\theta), \qquad
\xi_2 = - \xi_0 \sin(\theta) + \xi_m \cos(\theta)
\end{equation}
then by straightforward calculations we obtain
\begin{eqnarray}
\delta h_{1\mu\nu} &=& D_\mu \xi_{1\nu} + D_\nu \xi_{1\mu} +
\frac{a_{01}}{\cos(\theta)} [ \xi_1{}^\alpha D_\alpha h_{1\mu\nu} +
D_{(\mu} \xi_1{}^\alpha h_{1\nu)\alpha} ] \nonumber \\
\delta h_{2\mu\nu} &=& D_\mu \xi_{2\nu} + D_\nu \xi_{2\mu} -
\frac{a_{01}}{\sin(\theta)} [ \xi_2{}^\alpha D_\alpha h_{2\mu\nu} +
D_{(\mu} \xi_2{}^\alpha h_{2\nu)\alpha} ]
\end{eqnarray}
provided
\begin{equation}
\tan(2\theta) = - \frac{2a_{01}}{a_{04}}
\end{equation}
Thus in terms of these new fields the parts of the Lagrangian with two
derivatives are completely separated and the two metrics of bigravity
can be written simply as:
\begin{eqnarray}
g_{\mu\nu} &=& \eta_{\mu\nu} + a_{01} [ h_{0\mu\nu} + \tan(\theta)
h_{m\mu\nu} ] \nonumber \\
f_{\mu\nu} &=& \eta_{\mu\nu} + a_{01} [ h_{0\mu\nu} - \cot(\theta)
h_{m\mu\nu} ]
\end{eqnarray}
As for the potential terms, collecting the results of previous two
sections we can write the most general form for the cubic part as:
\begin{eqnarray}
{\cal L}_{10} &=& a_{01} m^2 [ h_{0\mu\nu} ( h_{m\mu\alpha}
h_{m\alpha\nu} - h_m h_{m\mu\nu} ) - \frac{1}{4} h_0 ( h_{m\mu\nu}{}^2
- h_m{}^2 ) ] \nonumber \\
 && + \frac{a_{04} m^2}{4} [ (1+8b_3) h_{m\mu\nu} h_{m\nu\alpha}
h_{m\alpha\mu} - (1+12b_3) h_m h_{m\mu\nu}{}^2 + 4b_3 h_m{}^3 ]
\end{eqnarray}
Note that here we have changed the normalization of $b_3$, in
particular, its special value now $b_3 = \frac{1}{8}$.

\section{Three fields}

At last let us consider cubic vertex for three spin-2 fields with
different masses. This time we choose the following general form for
the terms with two derivatives:
$$
{\cal L}_{12} \sim h_1 h_2 D^2 h_3 \oplus h_1 D^2 h_2 h_3 \oplus h_1
D h_2 D h_3
$$
while potential terms will be written as follows:
\begin{equation}
{\cal L}_{10} = b_1 h_{1\mu\nu} h_{2\nu\alpha} h_{3\alpha\mu} + b_2
h_{1\mu\nu} h_{2\mu\nu} h_3 + b_3 h_{1\mu\nu} h_2 h_{3\mu\nu} + b_4
h_1 h_{2\mu\nu} h_{3\mu\nu} + b_5 h_1 h_2 h_3
\end{equation}

Again we require that there are all three vector constraints with
possible quadratic corrections:
$$
\Delta {\cal C}_{1\nu} \approx D^\mu \frac{\delta {\cal L}_1}{\delta
h_{1\mu\nu}}, \qquad \Delta {\cal C}_{2\nu} \approx D^\mu 
\frac{\delta {\cal L}_1}{\delta h_{2\mu\nu}}, \qquad \Delta 
{\cal C}_{3\nu} \approx D^\mu \frac{\delta {\cal L}_1}{\delta 
h_{3\mu\nu}}
$$
or schematically:
\begin{eqnarray*}
\Delta {\cal C}_{1V} = D \frac{\delta {\cal L}_1}{\delta h_1} + (D h_2
+ h_2 D) \frac{\delta {\cal L}_0}{\delta h_3} + (D h_3 + h_3 D)
\frac{\delta {\cal L}_0}{\delta h_2} \\
\Delta {\cal C}_{2V} = D \frac{\delta {\cal L}_1}{\delta h_2} + (D h_1
+ h_1 D) \frac{\delta {\cal L}_0}{\delta h_3}  + (D h_3 + h_3 D)
\frac{\delta {\cal L}_0}{\delta h_1} \\
\Delta {\cal C}_{3V} = D \frac{\delta {\cal L}_1}{\delta h_3} + (D h_1
+ h_1 D) \frac{\delta {\cal L}_0}{\delta h_2}  + (D h_2 + h_2 D)
\frac{\delta {\cal L}_0}{\delta h_1}
\end{eqnarray*}
In this case the general solution for ${\cal L}_{12}$ has sixteen free
parameters, but we have fifteen possible fields redefinitions:
\begin{eqnarray}
h_{1\mu\nu} &\Rightarrow& h_{1\mu\nu} + s_1 h_{2\alpha(\mu}
h_{3\nu)\alpha} + s_2 h_2 h_{3\mu\nu} + s_3 h_3 h_{2\mu\nu} + 
\eta_{\mu\nu} ( s_4 h_{2\alpha\beta} h_{3\alpha\beta} + s_5 h_2 h_3)
\nonumber \\
h_{2\mu\nu} &\Rightarrow& h_{2\mu\nu} + s_6 h_{1\alpha(\mu}
h_{3\nu)\alpha} + s_7 h_1 h_{3\mu\nu} + s_8 h_3 h_{1\mu\nu} + 
\eta_{\mu\nu} ( s_9 h_{1\alpha\beta} h_{3\alpha\beta} + s_{10} h_1
h_3) \\
h_{3\mu\nu} &\Rightarrow& h_{3\mu\nu} + s_{11} h_{1\alpha(\mu}
h_{2\nu)\alpha} + s_{12} h_1 h_{2\mu\nu} + s_{13} h_2 h_{1\mu\nu} + 
\eta_{\mu\nu} (s_{14} h_{1\alpha\beta} h_{2\alpha\beta} + s_{15} h_1
h_2 ) \nonumber
\end{eqnarray}
and hence only one non-trivial coupling constant. Using this freedom
we choose explicit form for the ${\cal L}_{12}$ that in the massless
case $m_1 = m_2 = m_3 = 0$ corresponds to invariance of the Lagrangian
under the following gauge transformations:
\begin{eqnarray}
\delta h_{1\mu\nu} &=& D_\mu \xi_{1\nu} + D_\nu \xi_{1\mu} + a_0 [
\xi_2{}^\alpha D_\alpha h_{3\mu\nu} + D_{(\mu} \xi_2{}^\alpha
h_{3\nu)\alpha} + \nonumber \\
 && + \xi_3{}^\alpha D_\alpha h_{2\mu\nu} + D_{(\mu} \xi_3{}^\alpha
h_{2\nu)\alpha} ] \nonumber \\
\delta h_{2\mu\nu} &=& D_\mu \xi_{2\nu} + D_\nu \xi_{2\mu} + a_0 [
\xi_1{}^\alpha D_\alpha h_{3\mu\nu} + D_{(\mu} \xi_1{}^\alpha
h_{3\nu)\alpha} +  \\
 && + \xi_3{}^\alpha D_\alpha h_{1\mu\nu} + D_{(\mu} \xi_3{}^\alpha
h_{1\nu)\alpha} ] \nonumber \\
\delta h_{3\mu\nu} &=& D_\mu \xi_{3\nu} + D_\nu \xi_{3\mu} + a_0 [
\xi_1{}^\alpha D_\alpha h_{2\mu\nu} + D_{(\mu} \xi_1{}^\alpha
h_{2\nu)\alpha} + \nonumber \\
 && + \xi_2{}^\alpha D_\alpha h_{1\mu\nu} + D_{(\mu} \xi_2{}^\alpha
h_{1\nu)\alpha} ] \nonumber
\end{eqnarray}

Similarly, we require existence of all three scalar constraints with
possible corrections:
\begin{eqnarray*}
\Delta {\cal C}_1 \approx (D^\mu D^\nu - \frac{m_1{}^2}{(d-2)}
\eta^{\mu\nu}) \frac{\delta {\cal L}_1}{\delta h_{1\mu\nu}} \\
\Delta {\cal C}_2 \approx (D^\mu D^\nu - \frac{m_2{}^2}{(d-2)}
\eta^{\mu\nu}) \frac{\delta {\cal L}_1}{\delta h_{2\mu\nu}} \\
\Delta {\cal C}_3 \approx (D^\mu D^\nu - \frac{m_3{}^2}{(d-2)}
\eta^{\mu\nu}) \frac{\delta {\cal L}_1}{\delta h_{3\mu\nu}} 
\end{eqnarray*}
or schematically:
\begin{eqnarray*}
\Delta {\cal C}_1 &=& (D D - \eta) \frac{{\cal L}_1}{\delta h_1} + 
(D^2h_2 + Dh_2 D + h_2 D^2 + h_2) \frac{{\cal L}_0}{\delta h_3} \\
 && + (D^2h_3 + Dh_3 D + h_3 D^2 + h_3) \frac{{\cal L}_0}{\delta h_2}
\\
\Delta {\cal C}_2 &=& (D D - \eta) \frac{{\cal L}_1}{\delta h_2} + 
(D^2h_1 + Dh_1 D + h_1 D^2 + h_1) \frac{{\cal L}_0}{\delta h_3}  \\
 && + (D^2h_3 + Dh_3 D + h_3 D^2 + h_3) \frac{{\cal L}_0}{\delta h_1}
\\
\Delta {\cal C}_3 &=& (D D - \eta) \frac{{\cal L}_1}{\delta h_3} + 
(D^2h_1 + Dh_1 D + h_1 D^2 + h_1) \frac{{\cal L}_0}{\delta h_2}  \\
 && + (D^2h_2 + Dh_2 D + h_2 D^2 + h_2) \frac{{\cal L}_0}{\delta h_1} 
\end{eqnarray*}

Here we again have a number of possibilities that we consider one by
one.

\noindent
{\bf Two fields are massless $m_1 = m_2 = 0$}. No solution except for
$m_3 = 0$.

\noindent
{\bf One field is massless $m_1 = 0$}. Solution exists for $m_2 = m_3$
only again in agreement with the results of \cite{Met12} and (upon
identification $h_2 = h_3$) corresponds to the case with two fields.
Scalar constraints are algebraic, partially massless limit exists.

\noindent
{\bf All three fields are massive}. General solution:
$$
b_1 = \frac{a_0(m_1{}^2 + m_2{}^2 + m_3{}^2) + 4b_5}{2}, \qquad
b_2 = - \frac{a_0m_3{}^2 + 2b_5}{2} 
$$
$$
b_3 = - \frac{a_0m_2{}^2 + 2b_5}{2}, \qquad
b_4 = - \frac{a_0m_1{}^2 + 2b_5}{2}
$$
with $b_5$ as a free parameter. For general values of this parameter
all three scalar constraints contain terms with derivatives which are
absent only for 
\begin{equation}
b_5 = \frac{a_0(m_1{}^2 + m_2{}^2 + m_3{}^2)}{4}
\end{equation}
In this
$$
b_1 = a_0 ( m_1{}^2 + m_2{}^2 + m_3{}^2), \qquad
b_2 = - \frac{a_0 (m_1{}^2 + m_2{}^2 + 3m_3{}^2)}{4} 
$$
$$
b_3 = - \frac{a_0 (m_1{}^2 + 3m_2{}^2 + m_3{}^2)}{4}, \qquad
b_4 = - \frac{a_0 (3m_1{}^2 + m_2{}^2 + m_3{}^2)}{4}
$$

\noindent
{\bf One field partially massless $m_1{}^2 = (d-2)\kappa$}. There is a
solution provided the following relation holds:
\begin{equation}
(m_2{}^2 - m_3{}^2)^2 + 2 (m_2{}^2 + m_3{}^2) \kappa = d(d-2) \kappa^2
\label{eq}
\end{equation}
In this, scalar constraints for both massive fields are algebraic and
the Lagrangian is invariant under the following gauge transformations:
\begin{eqnarray}
\delta h_{1\mu\nu} &=& (D_\mu D_\nu - \kappa \eta_{\mu\nu} ) \xi
\nonumber \\
\delta h_{2\mu\nu} &=& \frac{a_0}{4} [
\frac{m_2{}^2+m_3{}^2-\kappa(d-2)}{m_2{}^2} h_{3(\mu}{}^\alpha
D_{\nu)} D_\alpha \xi + \frac{m_3{}^2-m_2{}^2-\kappa(d-2)}{m_2{}^2}
D_{(\mu} h_{3\nu)}{}^\alpha D_\alpha \xi \nonumber \\
 && + D^\alpha h_{3\mu\nu} D_\alpha \xi + 
(m_3{}^2-m_2{}^2+\kappa(d-6)) h_{3\mu\nu} \xi ] \\
\delta h_{3\mu\nu} &=& \frac{a_0}{4} [
\frac{m_2{}^2+m_3{}^2-\kappa(d-2)}{m_3{}^2} h_{2(\mu}{}^\alpha
D_{\nu)} D_\alpha \xi + \frac{m_2{}^2-m_3{}^2-\kappa(d-2)}{m_3{}^2}
D_{(\mu} h_{2\nu)}{}^\alpha D_\alpha \xi \nonumber \\
 && + D^\alpha h_{2\mu\nu} D_\alpha \xi + 
(m_2{}^2-m_3{}^2+\kappa(d-6)) h_{2\mu\nu} \xi ] \nonumber
\end{eqnarray}
There are two particular solutions of the relation (\ref{eq}) that
correspond to some two field cases given above.
\begin{itemize}
\item If masses of the second and third fields are equal $m_2 = m_3$
then this relation gives
$$
m_2{}^2 = m_3{}^2 = \frac{d(d-2)\kappa}{4}
$$
and this corresponds to the two field case where first field is
partially massless.
\item If the second field is also partially massless $m_2{}^2 =
(d-2)\kappa$ then we obtain
$$
m_3{}^2 = 2(d-3)\kappa
$$
exactly as in the two field case where the second field is partially
massless.
\end{itemize}

\section*{Conclusion}

In this paper we provide a complete (we hope) list of spin-2 cubic
interaction vertices with two derivatives. To include not only massive
and massless cases but also partially massless ones we work in (anti)
de Sitter space with dimension $d \ge 4$ and arbitrary value of
cosmological constant. To be as model independent as possible we work
with a simple metric formalism without any auxiliary or Stueckelberg
fields and use "brute force" method, i.e. we generate the most general
form of cubic vertex and require that Lagrangen equations produce
necessary constraints and/or lead to the invariance under some local
gauge transformations.

For the massive cases we have seen that one obtains solutions with one
free parameter but for the general values of this parameter the scalar
constraints contain dangerous derivative terms that lead to the
problem with causality. In all cases we considered there is a special
value of this parameter when these terms are absent. Moreover, it is
these special solutions that in de Sitter space admit partially
massless limit. Thus the problem of causality and the existence of
partially massless limit in massive (bi)gravity theories seem to be
closely related.

For all cases where partially massless field where present we provide
explicit form of the appropriate local gauge transformations with
scalar parameter.\footnote{Let us stress once again that it was
important that besides these gauge symmetry we still have appropriate
vector constraint.} Their rather complicated form suggests that they
may arise from gauge invariant Stueckelberg formalism as a result of
partial gauge fixing. Work in progress in this direction.

\vskip 1cm \noindent
{\bf Acknowledgment} Author is grateful to R.~R.~Metsaev and
E.~D.~Skvortsov for discussions and correspondence. The work was
supported in parts by RFBR grant No.11-02-00814.


\begin{thebibliography}{10}

\bibitem{Vas01}
M.~A. Vasiliev
{\it "Cubic Interactions of Bosonic Higher Spin Gauge Fields in
$AdS_5$",}
Nucl.Phys. {\bf B616} (2001) 106-162; Erratum-ibid. B652 (2003) 407,
  arXiv:hep-th/0106200.

\bibitem{AV02}
K.~B. Alkalaev, M.~A. Vasiliev
{\it "N=1 Supersymmetric Theory of Higher Spin Gauge Fields in AdS(5)
at the
  Cubic Level",}
Nucl.Phys. {\bf B655} (2003) 57-92, arXiv:hep-th/0206068.

\bibitem{Alk10}
K.B. Alkalaev
{\it "FV-type action for AdS(5) mixed-symmetry fields",}
JHEP {\bf 1103} (2011) 031, arXiv:1011.6109.

\bibitem{Vas11}
M.~Vasiliev
{\it "Cubic Vertices for Symmetric Higher-Spin Gauge Fields in
$(A)dS_d$",}
Nucl. Phys. {\bf B862} (2012) 341, arXiv:1108.5921.

\bibitem{Met05}
R.~R. Metsaev
{\it "Cubic interaction vertices of massive and massless higher spin
fields",}
Nucl. Phys. {\bf B759} (2006) 147, arXiv:hep-th/0512342.

\bibitem{Met06a}
R.~R. Metsaev
{\it "Gravitational and higher-derivative interactions of massive spin
5/2
  field in (A)dS space",}
Phys. Rev. {\bf D77} (2008) 025032, arXiv:hep-th/0612279.

\bibitem{Met07b}
R.R. Metsaev
{\it "Cubic interaction vertices for fermionic and bosonic arbitrary
spin
  fields",}
Nucl. Phys. {\bf B859} (2012) 13, arXiv:0712.3526.

\bibitem{Met12}
R.~R. Metsaev
{\it "BRST-BV approach to cubic interaction vertices for massive and
massless
  higher-spin fields",} arXiv:1205.3131.

\bibitem{BL06}
N.~Boulanger, S.~Leclercq
{\it "Consistent couplings between spin-2 and spin-3 massless
fields",}
JHEP {\bf 0611} (2006) 034, arXiv:hep-th/0609221.

\bibitem{BLS08}
N.~Boulanger, S.~Leclercq, P.~Sundell
{\it "On The Uniqueness of Minimal Coupling in Higher-Spin Gauge
Theory",}
JHEP {\bf 0808} (2008) 056, arXiv:0805.2764.

\bibitem{BSZ11}
Nicolas Boulanger, E.~D. Skvortsov, Yu.~M. Zinoviev
{\it "Gravitational cubic interactions for a simple mixed-symmetry
gauge field
  in AdS and flat backgrounds",}
J. Phys. {\bf A44} (2011) 415403, arXiv:1107.1872.

\bibitem{BS11}
Nicolas Boulanger, E.~D. Skvortsov
{\it "Higher-spin algebras and cubic interactions for simple
mixed-symmetry
  fields in AdS spacetime",}
JHEP {\bf 1109} (2011) 063, arXiv:1107.5028.

\bibitem{BPS12}
N.~Boulanger, D.~Ponomarev, E.D. Skvortsov
{\it "Non-abelian cubic vertices for higher-spin fields in anti-de
Sitter
  space",} arXiv:1211.6979.

\bibitem{Zin06}
Yu.~M. Zinoviev
{\it "On massive spin 2 interactions",}
Nucl. Phys. {\bf B770} (2007) 83-106, arXiv:hep-th/0609170.

\bibitem{Zin08}
Yu.~M. Zinoviev
{\it "On spin 3 interacting with gravity",}
Class. Quantum Grav. {\bf 26} (2009) 035022, arXiv:0805.2226.

\bibitem{Zin09}
Yu.~M. Zinoviev
{\it "On massive spin 2 electromagnetic interactions",}
Nucl. Phys. {\bf B821} (2009) 431-451, arXiv:0901.3462.

\bibitem{Zin10}
Yu.~M. Zinoviev
{\it "Spin 3 cubic vertices in a frame-like formalism",}
JHEP {\bf 08} (2010) 084, arXiv:1007.0158.

\bibitem{Zin10a}
Yu.~M. Zinoviev
{\it "On electromagnetic interactions for massive mixed symmetry
field",}
JHEP {\bf 03} (2011) 082, arXiv:1012.2706.

\bibitem{Zin11}
Yu.~M. Zinoviev
{\it "Gravitational cubic interactions for a massive mixed symmetry
gauge
  field",}
Class. Quantum Grav. {\bf 29} (2012) 015013, arXiv:1107.3222.

\bibitem{BSZ12}
I.~L. Buchbinder, T.~V. Snegirev, Yu.~M. Zinoviev
{\it "Cubic interaction vertex of higher-spin fields with external
  electromagnetic field",} arXiv:1204.2341.

\bibitem{Zin12}
Yu.~M. Zinoviev
{\it "On massive gravity and bigravity in three dimensions",}
arXiv:1205.6892.

\bibitem{BSZ12b}
I.~L. Buchbinder, T.~V. Snegirev, Yu.~M. Zinoviev
{\it "On gravitational interactions for massive higher spins in
$AdS_3$",}
  arXiv:1208.0183.

\bibitem{MMR09}
R.~Manvelyan, K.~Mkrtchyan, W.~Ruehl
{\it "Off-shell construction of some trilinear higher spin gauge field
  interactions",}
Nucl. Phys. {\bf B826} (2010) 1, arXiv:0903.0243.

\bibitem{MMR10a}
R.~Manvelyan, K.~Mkrtchyan, W.~Ruehl
{\it "General trilinear interaction for arbitrary even higher spin
gauge
  fields",}
Nucl. Phys. {\bf B836} (2010) 204, arXiv:1003.2877.

\bibitem{MMR10b}
R.~Manvelyan, K.~Mkrtchyan, W.~Ruehl
{\it "A generating function for the cubic interactions of higher spin
fields",}
Phys. Lett. {\bf B696} (2011) 410, arXiv:1009.1054.

\bibitem{MMR12}
R.~Manvelyan, R.~Mkrtchyan, W.~Ruehl
{\it "Radial Reduction and Cubic Interaction for Higher Spins in (A)dS
space",}
  arXiv:1210.7227.

\bibitem{GHR12}
G.~L. Gomez, M.~Henneaux, R.~Rahman
{\it "Higher-Spin Fermionic Gauge Fields and Their Electromagnetic
Coupling",}
JHEP {\bf 1208} (2012) 093, arXiv:1206.1048.

\bibitem{ST10}
A.~Sagnotti, M.~Taronna
{\it "String Lessons for Higher-Spin Interactions",}
Nucl. Phys. {\bf B842} (2011) 299, arXiv:1006.5242.

\bibitem{JT11}
E.~Joung, M.~Taronna
{\it "Cubic interactions of massless higher spins in (A)dS:
metric-like
  approach",}
Nucl. Phys. {\bf B861} (2012) 145, arXiv:1110.5918.

\bibitem{JLT12}
E.~Joung, L.~Lopez, M.~Taronna
{\it "On the cubic interactions of massive and partially-massless
higher spins
  in (A)dS",}
JHEP {\bf 07} (2012) 041, arXiv:1203.6578.

\bibitem{JLT12a}
E.~Joung, L.~Lopez, M.~Taronna
{\it "Solving the Noether procedure for cubic interactions of higher
spins in
  (A)dS",} arXiv:1207.5520.

\bibitem{JLT12b}
E.~Joung, L.~Lopez, M.~Taronna
{\it "Generating functions of (partially-)massless higher-spin cubic
  interactions",} arXiv:1211.5912.

\bibitem{Red}
REDUCE computer algebra system --- www.reduce-algebra.com.

\bibitem{RG10}
Claudia de~Rham, Gregory Gabadadze
{\it "Generalization of the Fierz-Pauli Action",}
Phys. Rev. {\bf D82} (2010) 044020, arXiv:1007.0443.

\bibitem{RGT10}
Claudia de~Rham, Gregory Gabadadze, Andrew~J. Tolley
{\it "Resummation of Massive Gravity",}
Phys. Rev. Lett. {\bf 106} (2011) 231101, arXiv:1011.1232.

\bibitem{HRS11}
S.~F. Hassan, Rachel~A. Rosen, Angnis Schmidt-May
{\it "Ghost-free Massive Gravity with a General Reference Metric",}
  arXiv:1109.3230.

\bibitem{HR11}
S.~F. Hassan, Rachel~A. Rosen
{\it "Bimetric Gravity from Ghost-free Massive Gravity",}
arXiv:1109.3515.

\bibitem{HR11b}
S.~F. Hassan, Rachel~A. Rosen
{\it "Confirmation of the Secondary Constraint and Absence of Ghost in
Massive
  Gravity and Bimetric Gravity",} arXiv:1111.2070.

\bibitem{HSS12b}
S.~F. Hassan, A.~Schmidt-May, M.~von Strauss
{\it "On Consistent Theories of Massive Spin-2 Fields Coupled to
Gravity",}
  arXiv:1208.1515.

\bibitem{HR12}
Kurt Hinterbichler, Rachel~A. Rosen
{\it "Interacting Spin-2 Fields",} arXiv:1203.5783.

\bibitem{HSS12a}
S.~F. Hassan, A.~Schmidt-May, M.~von Strauss
{\it "Metric Formulation of Ghost-Free Multivielbein Theory",}
arXiv:1204.5202.

\bibitem{RR12}
Claudia de~Rham, Sebastien Renaux-Petel
{\it "Massive Gravity on de Sitter and Unique Candidate for Partially
Massless
  Gravity",} arXiv:1206.3482.

\bibitem{HSS12}
S.~F. Hassan, Angnis Schmidt-May, Mikael von Strauss
{\it "On Partially Massless Bimetric Gravity",} arXiv:1208.1797.

\bibitem{HSS12d}
S.~F. Hassan, A.~Schmidt-May, M.~von Strauss
{\it "Bimetric Theory and Partial Masslessness with Lanczos-Lovelock
Terms in
  Arbitrary Dimensions",} arXiv:1212.4525.

\bibitem{DJW12}
S.~Deser, E.~Joung, A.~Waldron
{\it "Partial Masslessness and Conformal Gravity",} arXiv:1208.1307.

\bibitem{DSW13}
S.~Deser, M.~Sandora, A.~Waldron
{\it "Nonlinear Partially Massless from Massive Gravity?",}
arXiv:1301.5621.

\bibitem{RHRT13}
Claudia de~Rham, Kurt Hinterbichler, Rachel~A. Rosen, Andrew~J. Tolley
{\it "Evidence for and Obstructions to Non-Linear Partially Massless
Gravity",}
  arXiv:1302.0025.

\bibitem{DW01}
S.~Deser, A.~Waldron
{\it "Gauge Invariance and Phases of Massive Higher Spins in (A)dS",}
Phys. Rev. Lett. {\bf 87} (2001) 031601, arXiv:hep-th/0102166.

\bibitem{DW01a}
S.~Deser, A.~Waldron
{\it "Partial Masslessness of Higher Spins in (A)dS",}
Nucl. Phys. {\bf B607} (2001) 577, arXiv:hep-th/0103198.

\bibitem{DW01c}
S.~Deser, A.~Waldron
{\it "Null Propagation of Partially Massless Higher Spins in (A)dS and
  Cosmological Constant Speculations",}
Phys. Lett. {\bf B513} (2001) 137, arXiv:hep-th/0105181.

\bibitem{Zin01}
Yu.~M. Zinoviev
{\it "On Massive High Spin Particles in (A)dS",} arXiv:hep-th/0108192.

\bibitem{DW12}
S.~Deser, A.~Waldron
{\it "Acausality of Massive Gravity",} arXiv:1212.5835.

\bibitem{DJW13}
S.~Deser, E.~Joung, A.~Waldron
{\it "Gravitational- and Self- Coupling of Partially Massless Spin
2",}
Phys. Rev. {\bf D86} (2012) 104004, arXiv:1301.4181.

\bibitem{BDGH00}
N.~Boulanger, T.~Damour, L.~Gualtieri, M.~Henneaux
{\it "Inconsistency of interacting, multi-graviton theories",}
Nucl. Phys. {\bf B597} (2001) 127, arXiv:hep-th/0007220.

\end{thebibliography}
\end{document}